\documentclass[12pt,preprint]{aastex}

\newcommand{\rhod}{\rho_{\mathrm d}}
\newcommand{\nd}{n_{\mathrm d}}
\newcommand{\vd}{\mathbf{v}_{\mathrm d}}
\newcommand{\vecv}{\mathbf{v}}
\newcommand{\fdrag}{\mathbf{f}_{\rm d}}
\newcommand{\Msoy}{{\rm M}_\odot/{\rm yr}}

\shorttitle{Dust distribution in a stellar bowshock}
\shortauthors{van Marle et al.}

\begin{document}


\title{Computing the dust distribution in the bowshock of a fast moving, evolved star}


\author{A.J. van Marle, Z. Meliani and R. Keppens}
\affil{Centre for Plasma Astrophysics, K.U.Leuven,
    Celestijnenlaan 200B, Heverlee, B-3001, Belgium}

\and
\author{L. Decin}
\affil{Institute of Astronomy, K.U.Leuven, Celestijnenlaan 200D, Heverlee, B-3001, Belgium}


\begin{abstract}
We study the hydrodynamical behavior occurring in the turbulent interaction zone of a fast moving red supergiant star, where the circumstellar and  interstellar material collide. 
In this wind-interstellar medium collision, the familiar bow shock, contact discontinuity, and wind termination shock morphology forms, with localized instability development. 

Our model includes a detailed treatment of dust grains in the stellar wind, and takes into account the drag forces between dust and gas. The dust is treated as pressureless gas components binned per grainsize, for which we use ten representative grainsize bins. 
Our simulations allow to deduce how dust grains of varying sizes become distributed throughout the circumstellar medium.

We show that smaller dust grains (radius $<\,0.045\mu$m) tend to be strongly bound to the gas and therefore follow the gas density distribution closely, with intricate finestructure due to essentially hydrodynamical instabilities at the wind-related contact discontinuity. Larger grains which are more resistant to drag forces are shown to have their own unique dust distribution, with progressive deviations from the gas morphology. 
Specifically, small dust grains stay entirely within the zone bound by shocked wind material. The large grains are capable of leaving the shocked wind layer, and can penetrate into the shocked or even unshocked interstellar medium. 
Depending on how the number of dust grains varies with grainsize, this should leave a clear imprint in infrared observations of bowshocks of red supergiants and other evolved stars. 

\end{abstract}





\section{Introduction}
\label{sec-intro}
Hydrodynamical simulations of the circumstellar medium of low-to-intermediate mass stars (LIMS) have traditionally focused on the morphology of the gas and neglected the presence of dust grains in the stellar wind \citep[E.g.][]{wetal07}. 
Those simulations that do include dust tend to focus on the inner wind region \citep{w06} or deal with the proto-planetary disks surrounding very young stars \citep{pm06}. 

With recent advances in infrared observations, such as satellites like Spitzer and Herschel, it has become possible to make detailed infrared observations of the circumstellar environment. 
The higher sensitivity of these satellites has enabled us to observe the weak signatures of the bowshocks that form when the wind of a moving star collides with the surrounding interstellar medium (ISM). 
Since dust grains are the primary source of infrared radiation, it becomes necessary to fully integrate them into the simulations to determine how the distribution of dust grains correlates with the morphology of the circumstellar gas and how the presence of dust influences the structure of the shock that occurs where the stellar wind interacts with the ISM. 

As an example, we have investigated the behavior of dust grains in the circumstellar environment of a fast moving red supergiant star. 
As input parameters we use the values obtained for \object{$\alpha$-Orionis}, which has been observed in infrared with the AKARI Satellite  \citep{uetal08}. 
This star shows evidence of a bow shock, indicating that it is moving through the ISM at a  velocity of $40\,n_H^{-0.5}$km/s with $n_H$ the dimensionless local hydrogen density \citep{uetal08}. 

So far, most simulations of fast-moving stars have focused on hot, high mass stars \citep{be95,ck98,vmetal06}, while simulations of the bowshocks of cool, lower mass stars were done by \citet{wetal07}. 
However, none of these simulations took  the presence of dust into account. 

\section{Modeling gas and dust interactions}
\label{sec-hydro}
\subsection{Governing equations}If one wants to incorporate dustgrains in the hydrodynamical models, one should not only take into account the motion of the dustgrains themselves, it also becomes necessary to add the kinetic interaction between gas and dust to the equations of hydrodynamics. 

We use the {\tt MPI-AMRVAC} code \citep{metal07,keppens11}, which solves sets of near-conservation laws on an adaptive mesh. 
For this particular study, we added a new module to the code which calculates the behavior of dust grains as coupled to the familiar gas dynamic equations. 
The dust and gas are linked through drag forces which they exert on each other. 
Besides conservation of mass, the momentum and energy equations for the gas become
\begin{eqnarray}
\frac{\partial(\rho\vecv)}{\partial t}+\nabla\cdot(\rho\vecv\vecv)~&=&~-\nabla p+\sum_{d=1}^n\fdrag, \, \, \, \, \, \, \, \, \\
\frac{\partial e}{\partial t} +\nabla\cdot\biggl[(p+e)\vecv\biggr]~&=&~\sum_{d=1}^n \vecv\cdot\fdrag \\
\nonumber                                                                   &-&\frac{\rho^2}{m_{\rm h}^2}\Lambda(T)\,,
\end{eqnarray}
with $\rho$ the gas density, $\vecv$ the gas velocity, $p$ the thermal pressure and $e$ the total energy density for the gas. We assume a standard ideal gas law for closure. The right hand side source terms for the energy evolution describe optically thin radiative cooling and the work done by the drag forces $\fdrag$. Radiative losses depend on the hydrogen and electron particle densities (derived from $\rho$ assuming full ionization with hydrogen mass $m_{\rm h}$) and involve a temperature dependent cooling curve $\Lambda(T)$.  
As most of the gas in our simulation is at comparatively low temperature, we use a cooling curve $\Lambda(T)$ that extends down to 1~K, with tabulated info based on numerical calculations done with the {\tt cloudy} code \citep{fetal98}. 
The drag force for species ${\rm d}$ per unit volume $\fdrag$ is specified in Eq.~\ref{eq:drag}; since we treat multiple dust species we sum over the individual drag forces.

We follow the prescription from \citet{pm06}, treating the dust as a gas without pressure, since the internal energy of a dustgrain only influences its surface temperature, but has no influence on its movement. 
Therefore, the motion of each dust species can be treated with the equations of conservation of mass and momentum:
\begin{eqnarray}
\frac{\partial \rhod}{\partial t}+\nabla\cdot{(\rhod\vd)}~&=~0, \\
\frac{\partial (\rhod\vd)}{\partial t} +\nabla\cdot(\rhod\vd\vd)~&=~-\fdrag \,,
\end{eqnarray}
where $\rhod$ and $\vd$ are the mass density and velocity of the dust species. 
The drag force $\fdrag$ is given by a combination of Epstein's drag law for the subsonic regime and Stokes' drag law for the supersonic regime \citep{k75}, 
which in closed form writes as
\begin{equation}
\fdrag~=~-(1-\alpha)\pi \nd \rho a_{\rm d}^2 \Delta\,\vecv\sqrt{\Delta\,\vecv^2 + v_{\rm t}^2} \,.
\label{eq:drag}
\end{equation}
Its form involves the dust particle density $\nd$, the particle radius $a_d$ (assuming that the dust particles are spherical), the velocity difference between dust and gas $\Delta\,\vecv=\vecv-\vd$ and the thermal speed of the gas $v_{\rm t}=\tiny{\frac{3}{4}}\sqrt{3p/\rho}$ \citep{k75}. 
It also includes the sticking coefficient $\alpha$, which is a measure of the percentage of atoms that stick to the dust grain after collision. 
This treatment neglects interactions between the different dust species. 
However, because of the low particle densities ($\ll\,1$\,cm$^{-3}$) and the small collision cross-section ($<\,1\,\mu$m$^2$) of the individual grains, the gas-dust interaction will dominate over the interaction between dust particles. 
As in \citet{k75} we assume that $\alpha\,=\,0.25$, so 75\% of the collisions are supposed to be elastic. 


\subsection{Simulation parameters and setup}
For our simulation we use a 2D cylindrical grid with $r=0$ to 2~pc and $z=-2$ to $2$pc. The base grid has a resolution of 80$\times$160 gridpoints, and we allow grid adaptivity up to a maximum of six additional refinement levels, which translates in an effective resolution of 5\,120$\times$10\,240 gridpoints. 
The adaptive mesh is handled dynamically and traces the presence of dust, ensuring that the front of each expanding dust type is always fully resolved. 
We initialize the stellar wind by filling a spherical area of radius 0.1\,pc around the origin with wind material. 
Since the star is moving through the ISM, we let the ISM flow past the star at a constant velocity. 

As input parameters for our simulations we use the observations of \object{$\alpha$-Orionis}, a typical example of a fast moving evolved star, as obtained  by \citet{uetal08}. 
This gives us a gas mass-loss rate of $\dot{M}_{\rm gas}=3\times10^{-6}\Msoy$, injected at wind velocity $v_{\rm wind}=15$~km/s. We estimate the dust mass-loss rate to be $2.5\times10^{-3}\dot{M}_{\rm gas}$ \citep{vetal06}.
Since the winds of evolved LIMS are believed to be dust-driven \citep{k75}, which means that the gas is dragged along with the grains, we assume that the dust and gas in the free-streaming wind are closely coupled, so that the velocity difference vanishes.  
Therefore, we initialize the dust grains with the same terminal velocity as the gas. 
This wind collides with an assumed homogeneous ISM, which has a particle density of $n_{\rm ISM}=2$\,cm$^{-3}$ and, in the frame of reference of the star,  flows past the star with $v_{\rm ISM}=28.3$\,km/s.
For the dust particles we assume a minimum radius of $a_{\rm min}=0.005\,\mu$m and a maximum of $a_{\rm max}=0.25\,\mu$m \citep{detal06}, with particle density distributed over the grainsizes as $n(a)\sim\,a^{-3.5}$. We represent this distribution with particles of ten different sizes, taking $n=10$ in Eq.~1-2, logarithmically distributed over the total grainsize interval. Each dust species effectively represents all dust grains within part of an interval $\Delta\,a_{\rm d}$. The size of each grain type is given by 
\begin{eqnarray}
\label{eq:size}
\log{(a_{\rm d})}~&=&~\log{(a_{\rm min})} \\
\nonumber                  &+& \frac{{\rm d}-1}{10}\biggl[\log{(a_{\rm max})}-\log{(a_{\rm min})}\biggr].
\end{eqnarray}
We assume the dust to be composed of silicates, for which the internal particle density is 3.3~g/cm${^3}$ \citep{dl84}. Together with the mentioned size distribution, this then fixes the dust particle densities $\nd$, which in combination with the injection velocity (constant for all grain sizes) determine the dust mass-loss. 
The particle sizes and masses used in this paper can be found in Table~\ref{tbl-dust}.

\section{Gas and dust distributions}
\label{sec-results}
\begin{figure*}
\includegraphics[scale=.9]{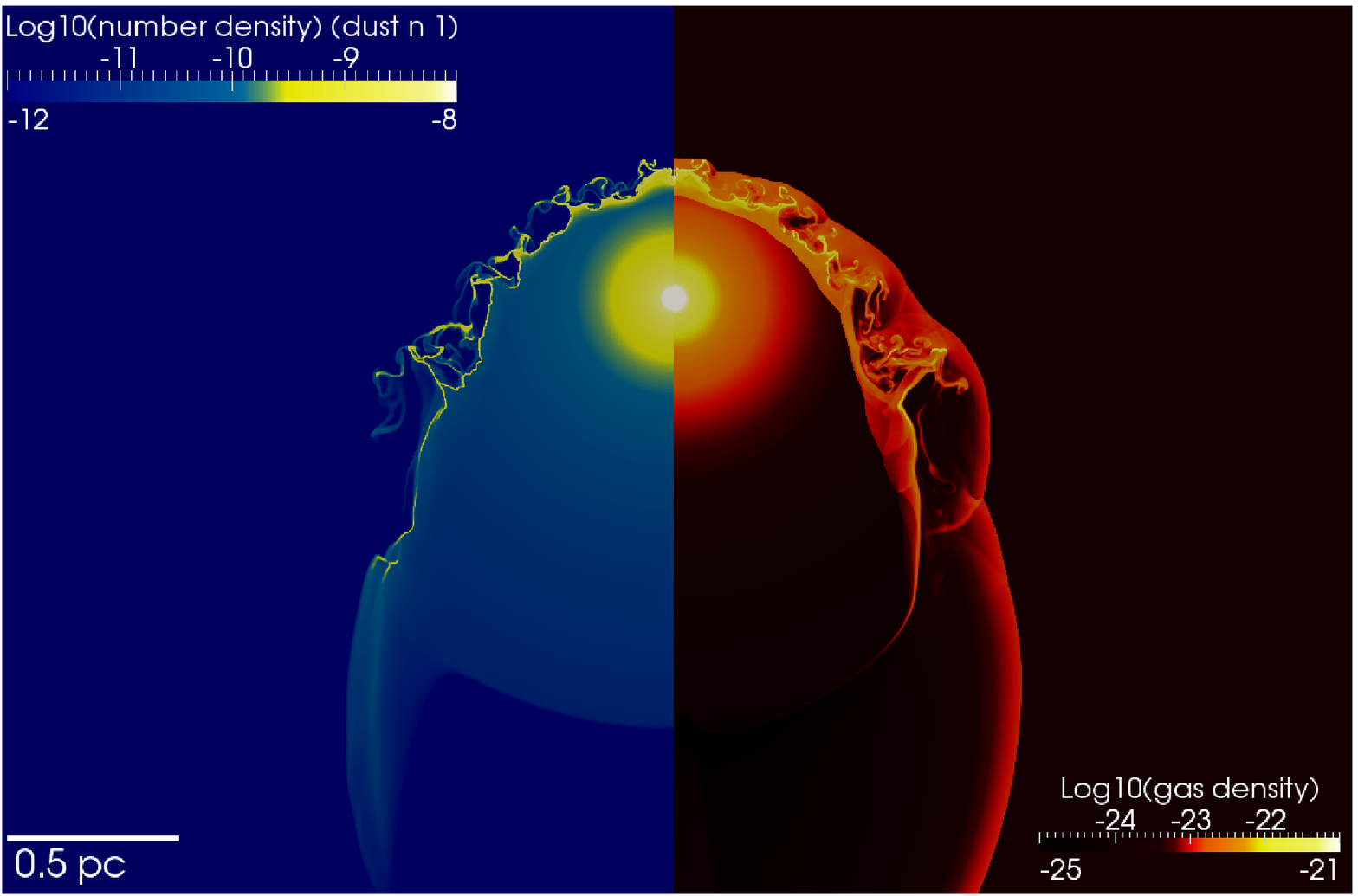}
\caption{This figure shows the gas density in g/cm$^3$(right) and particle density in \#/cm$^3$ of the smallest dust grains (type 1,\, radius 0.005 $\mu$m) in the circumstellar medium of a fast-moving red supergiant star after 75,000 years of simulation time. The general shape of the bowshock between the stellar wind and the interstellar medium (ISM) has reached an equilibrium position. The interaction region shows strong Rayleigh-Taylor instabilities as a result of the density difference between the shocked wind and the shocked ISM. The dust tends to pile up at the contact discontinuity and follows the local instabilities. This also indicates which part of the gas is wind material as these small grains tend to pile up at the transition between wind and ISM. 
\label{fig:rho_g_rho_d1}}
\end{figure*}

\begin{figure*}
 \includegraphics[scale=.9]{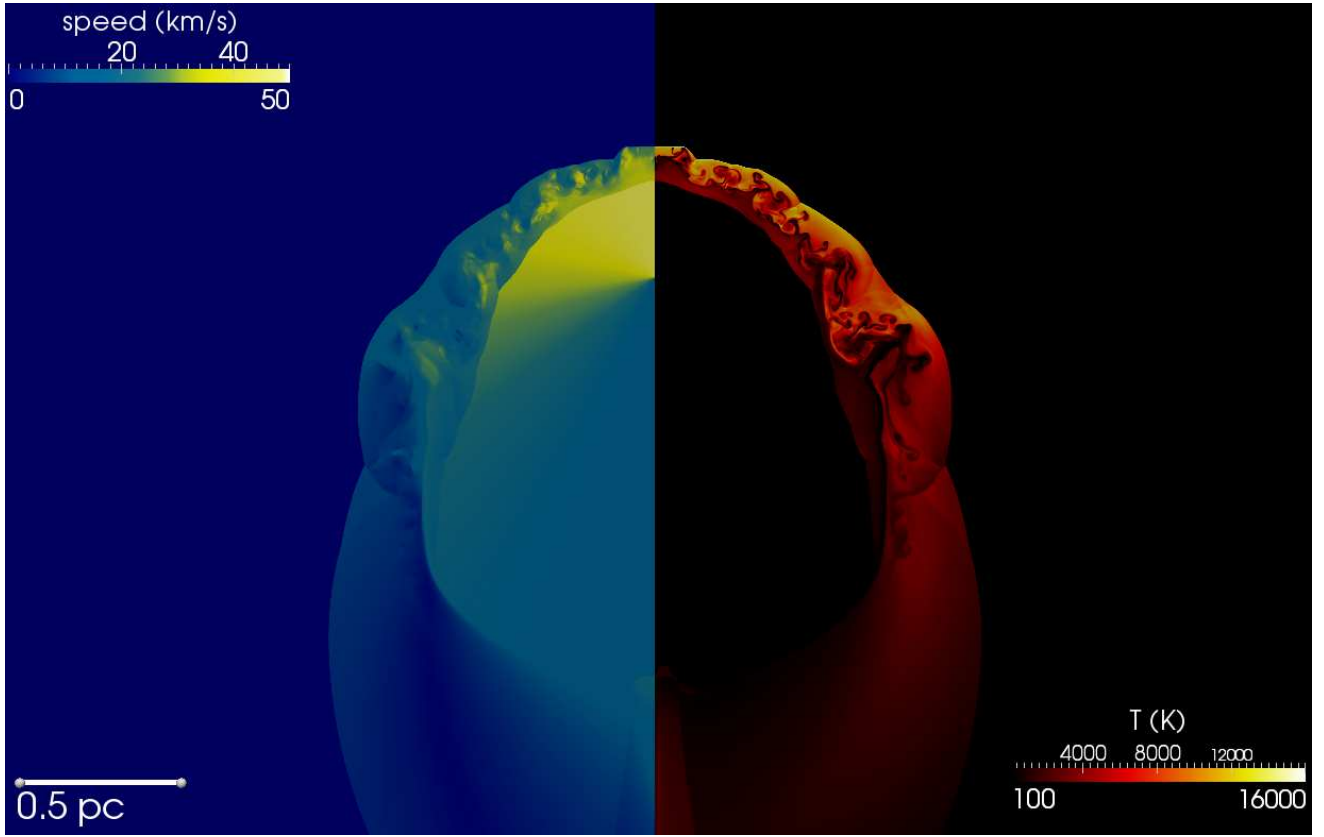}
\caption{The gas velocity in km/s (left) and temperature in K (right) at the same time as Fig.~\ref{fig:rho_g_rho_d1}. 
The shocked ISM is hotter than the shocked wind, since it starts with more kinetic energy and cools less efficiently due to its lower density. 
Note that the velocity is not in the co-moving frame, but rather in the rest-frame of the ISM.
\label{fig:vel_T}}
\end{figure*}

\begin{figure*}
\includegraphics[scale=.9]{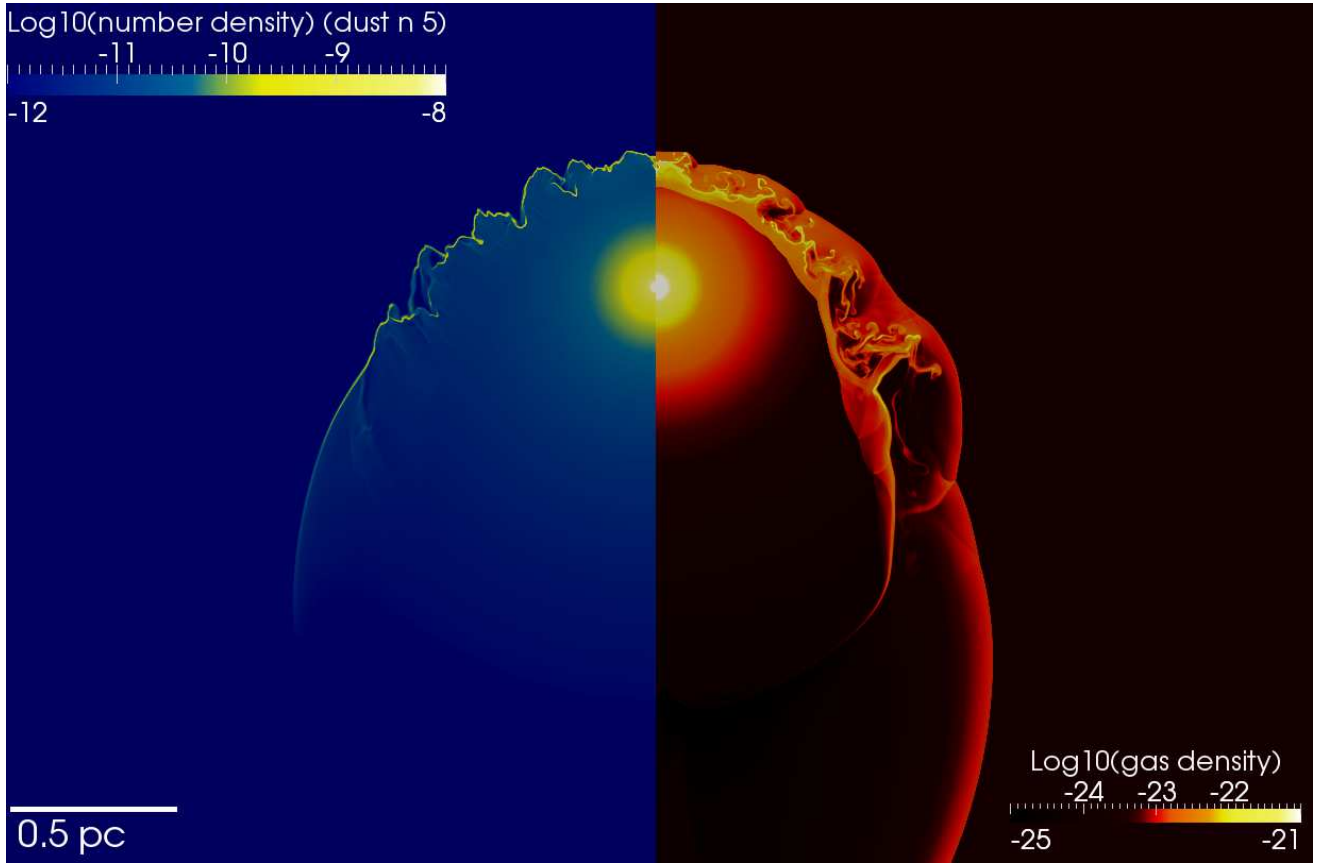}
\caption{Particle density in \#/cm$^3$ of the intermediate sized dust grains (type 5,\, $a$=0.0284$\mu$m) from Table~\ref{tbl-dust} on the left compared to the gas density (right) at the same time as Figs.~\ref{fig:rho_g_rho_d1} and \ref{fig:vel_T}. These grains tend to follow the forward shock, rather than the contact discontinuity, but still show the influence of the instabilities in the gas. 
\label{fig:rho_g_rho_d5}}
\end{figure*}

\begin{figure*}
\includegraphics[scale=.9]{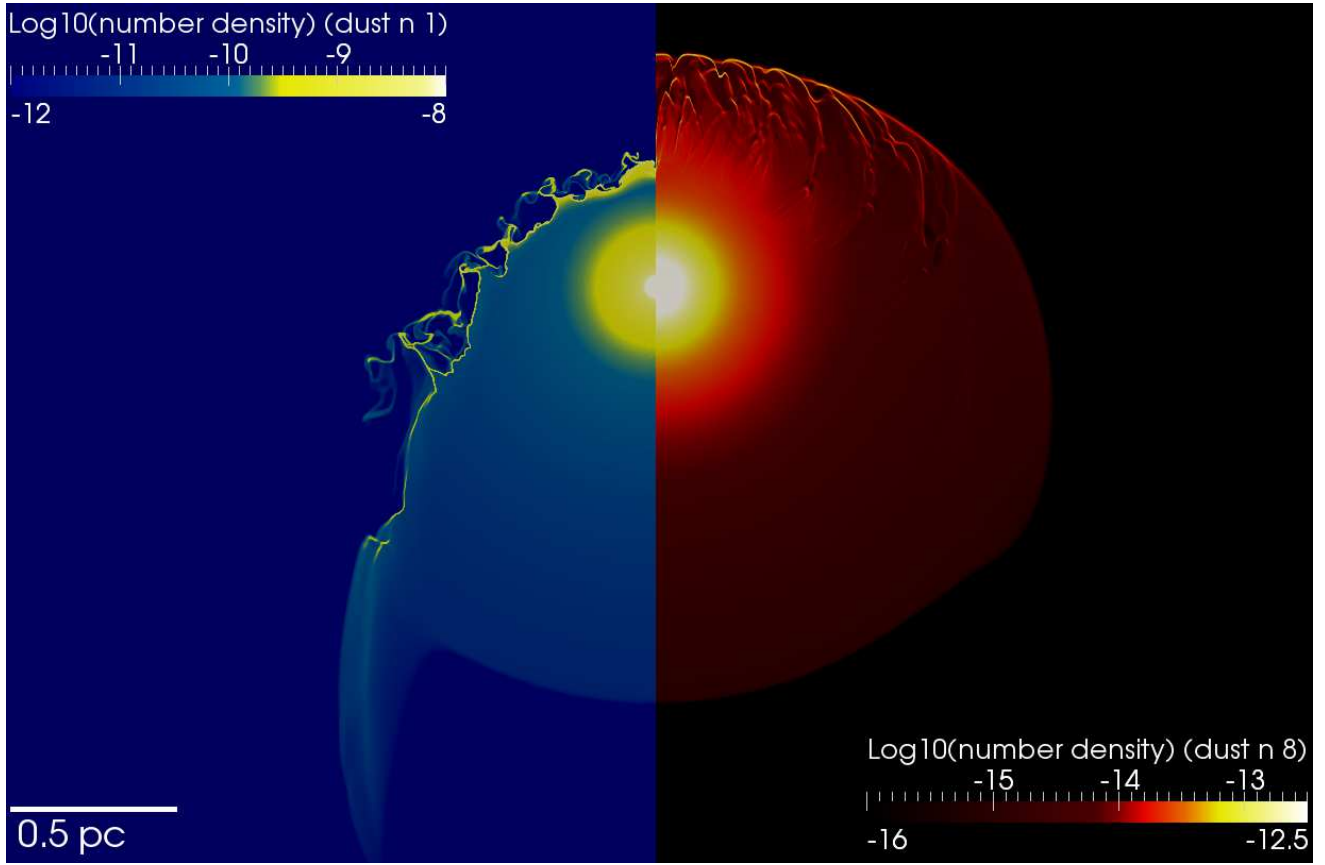}
\caption{The particle density distribution in \#/cm$^3$ of the larger dust grains (type 8,\, radius 0.105 $\mu$m) compared with the smallest grains (type 1,radius 0.005 $\mu$m), at the same time as Figs.~\ref{fig:rho_g_rho_d1}-\ref{fig:rho_g_rho_d5}. These grains show little influence from the shocked gas layer and penetrate deeply into the unshocked ISM before being stopped by the drag force.
\label{fig:rho_g_rho_d8}}
\end{figure*}

The results of our simulation are shown in Figs.~\ref{fig:rho_g_rho_d1} through \ref{fig:rho_g_rho_d8} after a period of 75,000~years, which show the gas density and the particle densities for the individual dust species. 

In Fig.~\ref{fig:rho_g_rho_d1} on the right we show the gas density, which clearly shows the bowshock and the wind termination shock, with the shocked gas layer in between. 
The distance between the shocks and the star is approximately 0.3\, pc, which corresponds to observations \citep{uetal08}. 
The shocked gas region shows that the contact discontinuity between shocked wind and shocked ISM is subject to Rayleigh-Taylor type instabilities, which dominate the interaction front. 
These instabilities result from the density difference between the shocked wind and ISM and are enhanced by the dust, which tend to move through the contact discontinuity. 
The temperature and absolute velocity of the gas are shown in Fig.~\ref{fig:vel_T}. 
The shocked wind initially has a temperature of about 4,000\,K. 
By the time it reaches the contact discontinuity, the temperature has decreased to about 1\,000\,K due to radiative cooling. 
This decrease in temperature leads to compression as the thermal pressure of the gas decreases, resulting in a thin, high density feature at the contact discontinuity. A more parametric study of this phenomenon has been shown in \citet{vmk11}.
The shocked ISM has a higher temperature ($\sim\,12,000$\,K) because of the higher relative velocity of the ISM relative to the star (about a factor 2) and cools less efficiently due to its lower density. 

In the free-streaming wind, the dust grains have the same velocity as the gas, so they feel no drag force. 
As the dust grains cross the shock, they begin to experience a strong drag force because of the sudden difference in velocity with the gas. 
Because the dust initially has the same velocity as the pre-shock gas and the shock reduces the gas velocity by a factor 4 (according to the Rankine-Hugoniot conditions for an ideal gas), the initial velocity difference is $\tiny{\frac{3}{4}}\,v_{\rm wind}$. 
The thermal velocity of the particles in the wind is of the same order of magnitude, so both the Stokes and Epstein drag forces (Eq.~\ref{eq:drag}) contribute to the total force. 
As the dust grains slow down due to the drag force, the Epstein regime  starts to dominate. 
How quickly the velocity difference decreases depends on the size of the dust grains. 
Large grains, which have a large momentum relative to their effective collision cross-section will take longer to slow down than small dust grains. 
As a result, the distribution of the dust grains over the gas becomes a function of the grain size.

The smallest dust grains (left hand side of Fig.~\ref{fig:rho_g_rho_d1}) tend to follow the same pattern as the gas. 
As the wind slows down due to the wind termination shock, so do the smaller dust grains, which tend to pile up in the shocked wind region. 
For the smallest grains, the coupling to the gas is so strong that they follow the Rayleigh-Taylor instabilities (Fig.~\ref{fig:rho_g_rho_d1}). 
Some cross the contact discontinuity to enter the shocked ISM and are carried away downstream. 
This plot also allows us to determine which part of the gas is wind material and which is interstellar: the small dustgrains stay in the wind material. 
This is shown clearly in the region behind the star, where the contact discontinuity is almost parallel with the motion of the ISM. 
Figure~\ref{fig:rho_g_rho_d5} shows the distribution of larger dust grains (type 5,\, $a_5$=0.0284$\mu$m), which have enough momentum to cross into the shocked ISM region and reach the forward shock. They are less susceptible to instabilities in the gas, though the effect of these instabilities is still visible.  

The largest dust grains can cross the shocked wind region due to their larger individual momentum and smaller collision surface relative to their mass. 
As a result, they mostly ignore the gas instabilities and eventually penetrate the forward bowshock to cross into the free-streaming (in the frame of the star) ISM. 
Since the free-streaming ISM is cold (assumed 1\,K for this simulation), the Stokes' regime dominates over Epstein's drag law in Eq.~\ref{eq:drag}. 
In the Stokes' regime, the drag force varies with the velocity difference squared. 
This makes the free-streaming ISM quite effective at stopping the oncoming dust grains (which move in the opposite direction, so the velocity difference is large) causing most of the particles to halt as soon as they cross into the free-streaming ISM (Fig.~\ref{fig:rho_g_rho_d5}).
Only the largest particles can continue. 
These will penetrate deeply into the ISM as they are all but immune to the drag force. 
This is shown in Fig.~\ref{fig:rho_g_rho_d8}, which shows the distribution of the type 8 dust grains (0.105$\mu$m) compared to the distribution of the smallest grains. 
The large grains almost completely ignore the morphology of the shocked gas shell and penetrate a considerable distance ($\sim\,0.1-0.2$\,pc) into the unshocked ISM. Since they are very low in number, they will not influence the gas morphology.


Since in nature the distribution of dust grains over the grain sizes is smooth, rather than existing of ten separate species, we can expect the distribution to be smoother than the relatively peaked positions found in our simulations. 
In the shocked gas, we could expect at least one, possibly several peaks in the grain density. 
The first peak would be in the shocked wind region, where the smallest grains will tend to pile up. 
A second peak may be visible just beyond the forward shock, where those grains that manage to cross the shocked gas region encounter the unshocked ISM. 
In the unshocked ISM, the dust grains would spread out according to their size, with the largest grains extending furthest. This would not lead to any distinctive features. The tendency of large grains to pass through the shocked gas, while small grains are trapped, will deliver a disproportionately large  number of large dust grains into the interstellar medium. 
For this particular simulation, the division lies by grains with a radius of approximately 0.045\,$\mu$m. Smaller grains will stay inside the shocked gas, larger grains tend to continue outward; the distance reached by the individual dust species are quantified in  Table \ref{tbl-dust}.

\section{Conclusions}
\label{sec-conc}
In the circumstellar nebula of a fast moving red supergiant, the small sized dust grains ($a<$0.045\,$\mu$m) tend to follow the motion of the gas. 
Therefore, observations of infrared radiation emitted by these particles will be representative of the gas morphology as well as the morphology of the dust itself. 
The larger grains tend to create a morphology of their own, as they are only weakly coupled to the gas. 

Because the small dust grains will tend to dominate the observations as a result of their larger number and larger collective radiative surface area, the infrared observations of circumstellar nebulae can be used to determine the location of the shocked gas directly. 
However, observers should be aware of this phenomenon as it may influence observational results. 
For example, depending on the distribution of the number of grains over the grain sizes, a single shock may be observed as a series of separate features due to dust grains piling up in different areas dependent on their size. 

In the future we plan to extend our simulations to a more general parameter study and attempt to duplicate the morphology of observed circumstellar nebulae around other cool stars, such as \object{CW Leo} \citep{letal10,c11}. 
We will also investigate in detail how the presence of dust influences the behavior of the gas.



\acknowledgments
A.J.v.M.\ acknowledges support from FWO, grant G.0277.08 and K.U.Leuven GOA/09/009 and the DEISA Consortium (www.deisa.eu), co-funded through the EU FP7 project RI-222919, for support within the DEISA Extreme Computing Initiative. 
Z. M. recognizes support from DART project: X201104616. 
Part of the simulations was done at the Flemish High Performance Computer Centre, VIC3 at K.U. Leuven. 
We thank Wang Ye at the Department of Physics \& Astronomy, University of Kentucky, for providing us with the radiative cooling curve.

\clearpage


\clearpage

\clearpage

\clearpage

\clearpage




\clearpage

\begin{table}
\begin{center}
\caption{Specification of the logarithmic distribution of dust grain sizes. Column 2 gives the size of each particle species, column 3 the size interval which the species represent (see Eq.\,\ref{eq:size}), column 4 gives their mass and the last column the maximum distance that they reach in the direction of the motion of the star before being stopped by the drag force.\label{tbl-dust}}
\begin{tabular}{crrrr}
\tableline\tableline
name & $a$[$\mu$m] & $\Delta\,a$[$\mu$m]  & mass per grain [g] & max. distance [pc]\\
\tableline
1 &  0.00500  &  0.00136 & 1.728$\times$10$^{-18}$ &   0.34 \\
2 &  0.00772  &  0.00346 & 6.366$\times$10$^{-18}$ &   0.35 \\
3 &  0.0119   &  0.00535 & 2.345$\times$10$^{-17}$ &   0.37 \\
4 &  0.0184   &  0.00826 & 8.639$\times$10$^{-17}$ &   0.37 \\
5 &  0.0284   &  0.0128  & 3.183$\times$10$^{-16}$ &   0.37 \\
6 &  0.0439   &  0.0197  & 1.172$\times$10$^{-15}$ &   0.38 \\
7 &  0.0679   &  0.0304  & 4.320$\times$10$^{-15}$ &   0.42 \\
8 &  0.105    &  0.0470  & 1.591$\times$10$^{-14}$ &   0.53 \\
9 &  0.162    &  0.0726  & 5.863$\times$10$^{-14}$ &   N/A\tablenotemark{a} \\
10 & 0.250    &  0.0441  & 2.160$\times$10$^{-13}$ &   N/A\tablenotemark{a} \\
\tableline
\end{tabular}
\tablenotetext{a}{This species has not been stopped by the drag force within the physical space of this simulation and continues to expand into the ISM}
\end{center}
\end{table}

\end{document}